# FAST QUERY-BY-EXAMPLE SPEECH SEARCH USING SEPARABLE MODEL

*Yuguang Yang[1], Yu Pan[2], Xin Dong[1], Minqiang Xu[1]*

[1]SpeakIn Technologies Co. Ltd.
[2]Beijing Institute of Technology, Beijing, China
{yangyuguang, dongxin, xuminqiang}@speakin.ai, panyu@bit.edu.cn

## ABSTRACT

Traditional Query-by-Example (QbE) speech search approaches usually use methods based on frame-level features, while state-of-the-art approaches tend to use models based on acoustic word embeddings (AWEs) to transform variable length audio signals into fixed length feature vector representations. However, these approaches cannot meet the requirements of the search quality as well as speed at the same time. In this paper, we propose a novel fast QbE speech search method based on separable models to fix this problem. First, a QbE speech search training framework is introduced. Second, we design a novel model inference scheme based on RepVGG which can efficiently improve the QbE search quality. Third, we modify and improve our QbE speech search model according to the proposed model inference scheme. Experiments on keywords dataset shows that our proposed method can improve the GPU Real-time Factor (RTF) from 1/150 to 1/2300 by just applying separable model scheme and outperforms other state-of-the-art methods.

## 1. INTRODUCTION

Query-by-example (QbE) speech search is the task of retrieving utterances relevant to a given spoken query [1], which has a very attractive prospect in many practical fields. Previous traditional approaches usually utilize methods based on text content matching [2,3], such as ASR which matches the text after decoding the spoken query. However, this approach cannot meet the requirements of speed and quality at the same time in practical applications. Thus, to avoid the decoding process of ASR, some methods [4-6] directly use the acoustic modeling part of ASR model to extract the features of audio signals, and then compare these features of different lengths by dynamic time wrapping (DTW) [7].

Recently, scholars have presented some acoustic word embedding (AWE) based methods for the usage of the low-resource speech scenarios. Unlike DTW, which uses frame-level variable length speech alignment modeling, acoustic word embedding (AWE) modeling refers to the extraction of speech audio into a fixed-length feature vector representation. In recent years, there has been a lot of work to validate the AWE modeling approaches over the DTW modeling approaches, especially in terms of reducing the computational effort.

Nevertheless, the AWE-based methods still have the problem of ignoring the efficiency of embedding extraction. To these ends, we propose a novel QbE speech search framework and model inference scheme to improve search accuracy and speed. Specifically, we want the model to be as parallelized as possible without losing performance benefits. Thus, the LSTM [8] based model is not the best choice. The structure similar to ResNet [9] has poor parallelization potential which will be time-consuming when searching long audio, but the powerful fitting effect like ResNet is indeed what we need. In the end, we chose the RepVGG [10] network because of its fitting capability, which is comparable to ResNet and has the advantage of excellent inference speed. Furthermore, on the basis of our inference scheme, we design a fast inference scheme that uses the separable model to improve the inference speed of the QbE speech search. We demonstrate the performance of our proposed method and the results show that the proposed method can improve the inference speed by at least ten times without reducing the retrieval effect by splitting the model and retraining part of parameters.

## 2. SYSTEM MODEL

Details of our proposed fast QbE speech search model will be introduced in this section, and its overall structure is shown in Fig.1.

Suppose we have N categories of training queries and M categories of inference queries, donated by $Q_{tr} = \{p_1, ..., p_N\}$ and $Q_{inf} = \{q_1, ..., q_M\}$ respectively. Besides there are $S_{tr} = \{r_1, ..., r_N\}$ and $S_{inf} = \{s_1, ..., s_M\}$ samples for each category. Thus, our dataset can be represented by $W_{tr} = \{p_1^1, p_1^2, ..., p_1^{r_1}, p_2^1, p_2^2, ..., p_2^{r_2}, ..., p_N^1, p_N^2, ..., p_N^{r_N}\}$ and $W_{inf} = \{q_1^1, q_1^2, ..., q_1^{s_1}, q_2^1, q_2^2, ..., q_2^{s_2}, ..., q_M^1, q_M^2, ..., q_M^{s_M}\}$, where all p and q stand for queries in fbank format. For each time, the model randomly selects a segment of filter-bank (fbank) in audio and predicts the word label when training.

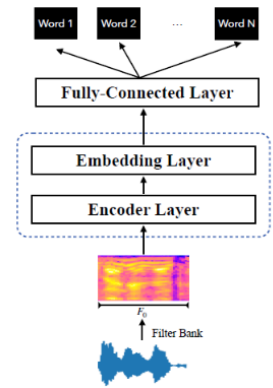

Figure 1: Overview of our proposed fast QbE speech search model.

## 2.1. Fast QbE speech search training framework

As shown in Fig.1, our proposed fast QbE speech search framework mainly consists of four parts, which are input representation, encoder layer, embedding layer and loss function, respectively.

### 2.1.1. Input representation
In our paper, every element that is cut out from fbank vectors of long audios should be padded to fixed frames $F_0$ for learning acoustic word embedding. To be specific, temporal context padding [14] is applied in our case due to the difference of word frame lengths.

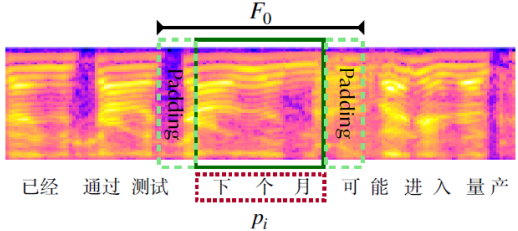

Figure 2: Illustration of the input representation.

### 2.1.2. Encoder layer
Suppose input representation $x$ is randomly selected from $W_{tr}$, it goes through the encoder to get output $y$:

$$y_{B,C,T^m,D_f^m} = enc(x_{B,T,D_f}) \quad (1)$$

where $B$ denotes the batch size, $T$ denotes the frame length of input fbank vector, which is equal to $F_0$ in the training phase, $D_f$ denotes dimensions of fbank, $T^m$ denotes the output middle hidden vector length on the dimension $T$, which will be $T_0^m$ in training phase, $D_f^m$ denotes the hidden vector length on $D_f$ dimension, $C$ denotes the output dimension of CNN operator.

The aim of this encoder layer is to embed the fbank features to higher-level representation, where TDNN [11]-BLSTM [8], SE-ResNet34 [12] as well as RepVGGA1 [10] are chosen in the experiment based on our previous experience on Voiceprint Recognition task.

On the basis of ResNet34[9], SEResNet [12] adds a Squeeze-and-Excitation (SE) structure which is applied to residual branches of each module of ResNet34.

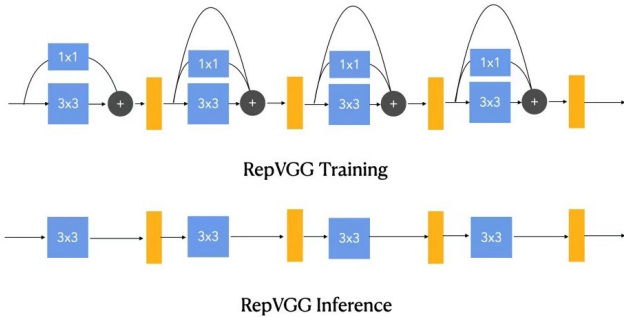

Figure 3: Difference between RepVGG training and inference.

However, SEResNet [12] is not quite efficient during the inference procedure due to its multi-branch structure. To have a better inference speed as well as a comparable result with SEResNet [12], we tried RepVGG [10] as our encoder, which is similar to ResNet [12]. RepVGG [10] only has $1 \times 1$, $3 \times 3$ convolution (conv) operation as well as short cut branch. For each block, $y_{blk} = x_{blk} + g(x_{blk}) + f(x_{blk})$, where $x_{blk}$ and $y_{blk}$ stand for input and output features of the block, $g(x)$ means the $1 \times 1$ conv branch and $f(x)$ means $3 \times 3$ branch. All the branches will be applied between non-linear layers, so it means that 3 branches of the block as well as its batch normalization (bn) layer could be represented as one single group of parameters without regarding the memory accuracy error. The overall structure of RepVGG [10] training and inference are shown in Fig. 4. Hence, it could obviously enhance the inference speed with almost no loss of quality.

### 2.1.3. Embedding layer
After we get a high dimensional representation of the audio signal $y$, it still needs an embedding layer to get a fixed length vector $\vec{e}$ for downstream inference:

$$\vec{e} = W\left(pool\left(y_{B,C,T^m,D_f^m}\right)\right) + b \quad (2)$$

To aggregate information of the total $T$ frames to a single vector representation, a pooling layer is applied, which is a concept from speaker recognition. Different from the traditional pooling layer, which is a simple combination of mean and std value, here we use the multi-head attention pooling layer [13]. For the sequence of hidden states $H_s = \{h_s^1, h_s^2, ..., h_s^n\}$, we generate a scalar weight for each sequence, and apply the weights with softmax output to mean and std value of the sequence. The multi-head method is set by splitting the sequence into $N_h$ sub-vectors on the hidden sequence, where $N_h$ means the number of heads.

After the pooling layer, a simple linear layer is applied for projecting the hidden vector to a fixed dimension $D_0$.

### 2.1.4. Loss functions
Additive margin softmax (AMsoftmax [14]) loss is proposed to match the cosine similarity method which is used to compare the word vectors. AMsoftmax [14] loss aims to expand the inter-class distance of word vectors and reduce their inner-class distance. Specifically, the AMsoftmax [14] loss is defined as:

$$L_{AMS} = -\frac{1}{n}\sum_{i=1}^{n} log \frac{e^{s \cdot (W_{y_i}^T f_i - m)}}{e^{s \cdot (W_{y_i}^T f_i - m)} + \sum_{j=1, j \neq y_i}^{c} e^{s W_j^T f_i}} \quad (3)$$

where f is the input of the last fully connected layer ($f_i$ is the i-th sample), $W_j$ is the j-th column of the last fully connected layer, C is the number of categories, m is the margin between different classes, s is a scaling factor, and we set m and s as 0.2 and 30 in our case. More detailed derivation and definition of the formula can be referred to [14,15].

Moreover, a penalization [16] P is added to loss function in order to lower the correlation of different heads, which is defined as:

$$P = \|AA^T - I\|_2^F \quad (4)$$

where A represents the weight matrix in the pooling layer, $\|\cdot\|_2^F$ stands for Frobenius norm. More detailed information can be referred to [16].

## 3. NOVEL INFERENCE SCHEME

### 3.1. Basic inference framework

The basic inference method with cosine similarity aims to retrieve the sequence with the method like sliding window, which ignores the correlations between similar words. Thus, a competition mechanism between words is introduced in our models.

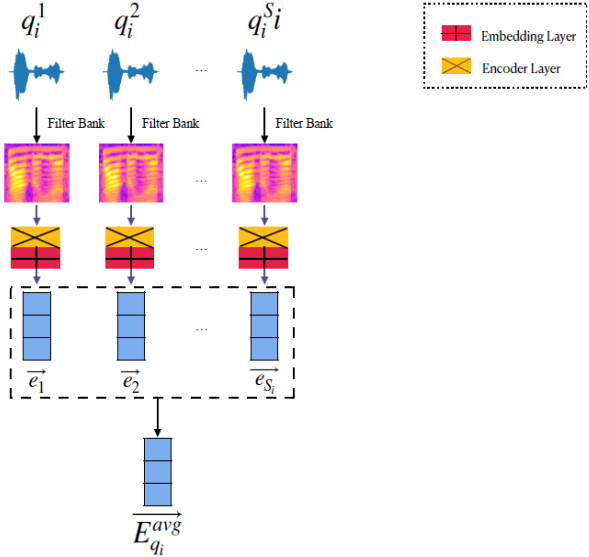

Figure 4: Average query vectors of the same word.

For certain $q_i$ from $Q_{inf}$, which has $s_i$ samples. We can put $s_i$ samples into model, then embedding $E_{q_i} = \{\vec{e}_1, \ldots, \vec{e}_{s_i}\}$ can be got. Besides, temporal context padding [16] will be used if the frame length is shorter than $F_0$. An average operation [4] is applied on $E_{q_i}$ to calculate $\vec{E}_{q_i}^{avg}$ for each word, in order to search in the next phase:

$$\vec{E}_{q_i}^{avg}: avg(E_{q_i}) = \frac{1}{s_i}\sum_{j=1}^{s_i}\frac{\vec{e}_j}{\|\vec{e}_j\|} \quad (5)$$

We can get a new query set $V_{inf} = \{\vec{E}_{q_1}^{avg}, \vec{E}_{q_2}^{avg}, \ldots, \vec{E}_{q_M}^{avg}\}$.

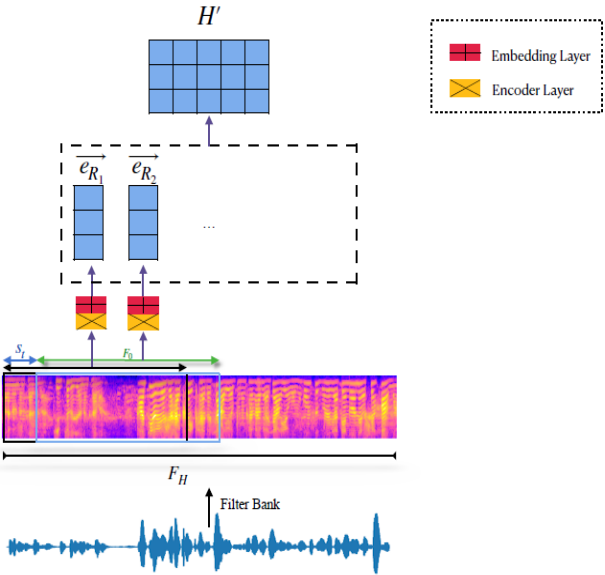

Figure 5: Basic inference schematic.

Suppose we got testing audio in the form of fbank vector called $H$, whose frame length is $F_H$. Let the input frame series be $H = \{h_1, h_2, \ldots, h_{F_H}\}$, a sliding window method is applied on $H$. Sliced fbank vector $S_v$ could be represented as:

$$\{R_1 = [h_1, h_2, \ldots, h_{F_0}],$$
$$R_2 = [h_{1+s_t}, h_{2+s_t}, \ldots, h_{F_0+s_t}], \ldots, \quad (6)$$
$$R_{\frac{F_H-F_0}{s_t}+1} = [h_{F_H-F_0+1}, h_{F_H-F_0+2}, \ldots, h_{F_H}]\}$$

where $s_t$ stands for stride of inference phase. After getting through well-trained model, output of middle feature map will be $H' = \{\vec{e}_{R_1}, \vec{e}_{R_2}, \ldots, \vec{e}_{R_{\frac{F_H-F_0}{s_t}+1}}\}$. For each $\vec{e}$ in $H'$, we calculate cosine similarity with each element in $V_{inf}$, each value in cosine matrix will be:

$$C_{sim} = 0.5 * \cos(\vec{e}_{R_i}, \vec{E}_{q_i}^{avg}) + 0.5 \quad (7)$$

where $\vec{e}_{R_i}$ and $\vec{E}_{q_i}^{avg}$ stands for embedding of audio predictions and queries, and $C_{sim}\epsilon[0,1]$. A similarity score matrix will be generated named $C_{mat}$ with shape $[M, \frac{F_H-F_0}{s_t}+1]$.

### 3.2. Competitive mechanism

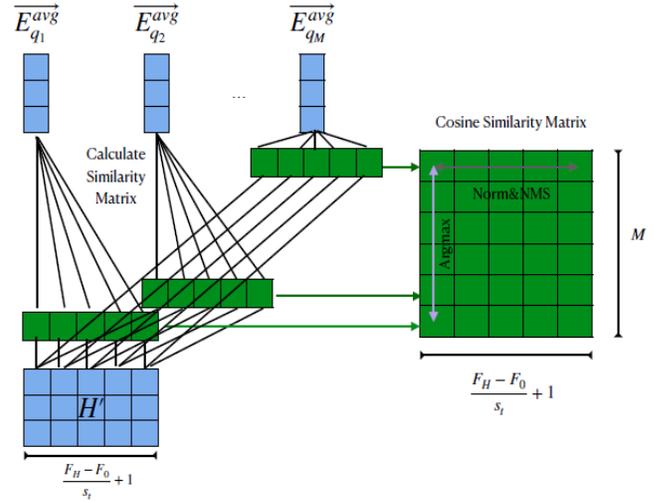

Figure 6: Calculate score matrix.

Briefly applying a threshold to the cosine similarity matrix $C_{mat}$ will get activated queries as well as its position, and we pick the word with the largest probability among these activated values as our target.

However, the method above still has its flaws. First, it does not consider the bias between the test audio and templates. So, for a certain template, it may always get a higher value than others, which will overwhelm other queries. In order to solve it, we apply T-norm [17] like operation on the dimension of $T$ to impair this bias.

Moreover, when the time of duration for a certain word is too long or some words may have overlaps, it will cause multiple detection which is similar with object detection task. Let a cosine similarity matrix with shape $[M, \frac{F_H-F_0}{s_t}+1]$. For certain query $q_i$, $L_w^i = \{l_w^1, l_w^2, \ldots, l_w^{s_i}\}$ stands for the set of query lengths on fbank vectors. An average length for each word on middle feature map $F_{nms}^i = avg(L_w^i)/C_r$ will be calculated, where $C_r$ stands for the down sampling ratio of our encoder. Then suppress all cosine similarity values to zero except for the maximum one on frame dimension of score matrix within area $F_{nms}^i$. And the whole structure is shown in Fig.6. For more concepts and details, it can be referred to non-maximum suppression [18] in object detection tasks.

## 4. SEPARABLE MODEL

### 4.1. Fast inference framework

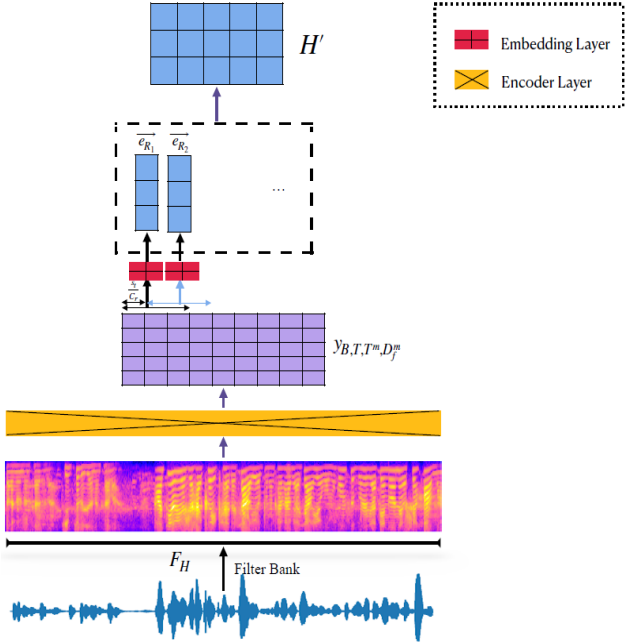

Figure 7: Fast inference schematic.

Previous inference approaches usually search the audio signal by sliding window on audio signals or fbank vectors. Suppose the stride time span is $s_t$ which is usually smaller than $F_0$, for each time we slide the window, there will be $F_0 - s_t$ length of the same information in fbank, which is quite time-consuming. In order to enhance the speed of model inference and information utilization, we separate the model into two parts when testing.

Suppose the input of model is $x_{B,T,D_f}$, the length of test audio fbank is $F_H$. We could make the calculation of encoder to get a hidden representation of $y_{B,C,T^m,D_f^m}$. Then, we could apply sliding window operation on the dimension $T^m$ of this hidden feature map, which slices length $T_0^m$ on $T^m$, and the new time span will be $s_t/C_r$. It is obvious that this scheme is feasible because of the translational equivalence of the convolution operation. Besides, it will dramatically improve the information utilization as well as inference speed. Based on our idea, the single-branch network should be applied. Obviously, the RepVGG [10] is a very good choice for its training in multi-branch and testing in the single-branch network.

### 4.2. Inconsistencies in inference calculations

In this section, we will elaborate on the inconsistencies in basic and fast inference architectures. Besides, we will explain why it is necessary to turn off the padding in the convolution operator.

First, we need to emphasize some basic concepts. Since the training of multi-branch network needs to make the output results of convolution kernel of different kernel size superimpose, the padding operation is required in the convolution operator and the padding operation is equivalent to zero paddings around the input of the convolution kernel. Second, due to the consistency of training and testing, the model structure of the basic scheme can be replaced with any network that fits the classification task well.

To clarify our computational steps, let's simplify our search framework. Suppose the input $H$ here is a sequence $\{h_1, h_2, h_3, h_4, h_5, h_6\}$, $F_0$ equals to 3, the encoder layer of our model here is only one layer conv operator with shape $1 \times 3$ and padding equals to 1. We define $C_{op}(\cdot)$ as the operation of convolution operator, $K_o(\cdot)$ as the operator of kernel calculation inside convolution operator, $concat(\cdot)$ as concatenation operator, $emb(\cdot)$ as the operation from $y_{B,C,T^m,D_f^m}$ to $\vec{e}$.

In our basic scheme of Fig.5, suppose $F_0$ here equals to 3 and $s_t$ here equals to 1. The calculation of base scheme will be like the follow. We slice the input $H$ into $S_v = \{R_1 = [h_1, h_2, h_3], R_2 = [h_2, h_3, h_4], ...\}$. Because of the padding parameter is 1, the convolution and embedding process of can be $R_1$ presented as:

$$C_{op}(R_1) = concat(K_o[0, h_1, h_2], K_o[h_1, h_2, h_3], K_o[h_2, h_3, 0]) \quad (8)$$

$$\vec{e}_{R_1} = emb\left(C_{op}(R_1)\right) \quad (9)$$

Then, the calculation of fast scheme in Fig.7, which puts long frame vector all into encoder in this case will be:

$$C_{op}(H) = concat(K_o[0, h_1, h_2], ..., K_o[h_2, h_3, h_4]) \quad (10)$$

Since padding operation will introduce nonlinear calculation on the $T$ dimension, resulting in our well-trained embedding layer ineffective in fast scheme. Properly speaking, we could not calculate same $\vec{e}_{R_i}$ as basic scheme under the condition of the fast scheme.

In the case of the convolution operator without the padding operation, the situation is different. The calculation output of $C_{op}(S_v)$ in the basic scheme is:

$$\begin{aligned} C_{op}(R_1) &= K_o([h_1, h_2, h_3]), \\ C_{op}(R_2) &= K_o([h_2, h_3, h_4]), \\ &... \end{aligned} \quad (11)$$

$$\begin{aligned} \vec{e}_{R_1} &= emb\left(C_{op}(R_1)\right) = K_o([h_1, h_2, h_3]), \\ \vec{e}_{R_2} &= emb\left(C_{op}(R_2)\right) = K_o([h_2, h_3, h_4]), \\ &... \end{aligned} \quad (12)$$

The calculation output in the fast scheme is $C_{op}(H) = concat([K_o([h_1, h_2, h_3]), K_o([h_2, h_3, h_4]), ...])$, and the $\vec{e}$ set could be calculated by putting sliced $C_{op}(H)$ vector into the embedding layer, which will be equal to results in the basic scheme.

### 4.3. Retrain embedding layer

Due to the difference between basic inference and fast inference, for a single convolution kernel, the edge of the convolution operator of basic inference are zeros, while the edge of the convolution operator of fast inference is context information.

Because of the inconsistency of the reception field of the CNN kernel, it could not achieve this by simply separate the RepVGG [10] model. To eliminate inconsistency, it needs to reload the pre-trained RepVGG-deploy [10] embedding parameters into a new structure which just changes the padding of conv operator to zero on the dimension of $T$ and retrain parameters of embedding layer only on the same task. Furthermore, it will still have a slight difference when shutting the padding. And $T^m$ will be smaller this time if the input still is the same size when finetunes our model. To use the encoder parameters, we propose two methods to achieve our aim.

Method1, enhance the length of $F_0$ when training, here we change the input frame length $F_0$ from 160 to 296 in our experiment.

Method2, keep the same input size $F_0$ as the first training. And retrain the parameters of the embedding layer. It can also enhance the number of our pooling layer here, which is related to the $T^m$ in our model. Since too few parameters in the embedding layer may degrade the effect of the model, we can use some small tricks to increase the number of parameters, such as increasing hidden size in LSTM [8].

## 5. EXPERIMENTS

### 5.1. Dataset

Because of the lack of the public dataset in this area, we choose the inner labeled ASR data in the company as our dataset. First, we calculate the word frequency in the data. Then, we choose 2000 words with its show-up frequency from up to down and clamp it to 2000, which means we randomly pick 2000 samples for one word if it has over 2000 samples inside our dataset. After that, we apply 9:1 ratio on the data as our train and valid dataset. Train and valid dataset consist of about 2,000,000 word-segments. Test set Ⅰ consists of 40 sorts of vocabulary (OOV) Mandarin keywords. Test set Ⅱ is chosen from our Cantonese data, which contains over 2k samples of 150 words. All of audios have been transferred to 60-dim fbank in our case.

### 5.2. Metrics

Because of the difference in our inference method and the purpose of enhancing the ability to evaluate the performance of usage scenarios, there are new metrics proposed.

First, the concept of overlap is defined as:

$$overlap = \begin{cases} True & abs\left(cent(Q_l) - cent(Q_p)\right) \leq t \\ False & abs\left(cent(Q_l) - cent(Q_p)\right) > t \end{cases} \quad (13)$$

where $cent(Q_l)$ and $cent(Q_p)$ means the center of labeled and predicted words.

Similarly, overlap ratio which means the proportion of the overlapping region to the labeled region is defined as:

$$overlap\ ratio = \frac{area(Q_l) \cap area(Q_p)}{area(Q_l)} \quad (14)$$

True Positive (TP) will be added only if there is the max overlap ratio, which means there will be only one when several predicted results overlap the same label word.

False Positive (FP) will be added when the prediction is not overlapped with the label, which means it will be added F if there are F predictions raised even for the same position on audios.

False Negative (FN) will be added when there is no prediction overlap certain label.

Mean Average Overlap (MAO) is defined as the mean average overlap ratio among labels.

### 5.3. Encoder

Table 1: Test results on Cantonese set of different encoders

| Testset Ⅱ | Threshold | Precision | Recall | F1 | MAO |
|---|---|---|---|---|---|
| 2k TDNN-BLSTM | 0.75 | 21.5 | 82.5 | 34.1 | **54.7** |
| | 0.8 | 35.3 | 60.2 | **44.5** | 40.2 |
| | 0.825 | 39.6 | 42.6 | 41.1 | 28.5 |
| | 0.85 | 39.9 | 24.4 | 30.3 | 16.5 |
| 2k SEResNet34 | 0.75 | 13.0 | 87.8 | 22.6 | **57.8** |
| | 0.8 | 28.3 | 71.2 | 40.5 | 47.1 |
| | 0.825 | 35.8 | 57.6 | **44.2** | 38.4 |
| | 0.85 | 40.9 | 38.0 | 39.4 | 25.8 |
| 2k RepVGGA1 | 0.75 | 11.9 | 90.2 | 21.1 | **59.3** |
| | 0.8 | 27.9 | 77.0 | 41.0 | 51.0 |
| | 0.825 | 35.0 | 62.8 | **45.0** | 41.6 |
| | 0.85 | 41.1 | 44.9 | 42.9 | 29.5 |

Table 1 gives results on the Cantonese testset, which is trained on Mandarin dataset. This indicates that the model has the potential of cross-lingual retrieval.

### 5.4. Penalization

Table 2: Test results on Mandarin with penalization

| Testset Ⅱ | Threshold | Precision | Recall | F1 | MAO |
|---|---|---|---|---|---|
| SEResNet34 | 0.75 | 21.5 | 82.5 | 34.1 | **54.7** |
| | 0.8 | 35.3 | 60.2 | **44.5** | 40.2 |
| | 0.825 | 39.6 | 42.6 | 41.1 | 28.5 |
| | 0.85 | 39.9 | 24.4 | 30.3 | 16.5 |
| SEResNet34 + P | 0.75 | 32.6 | 90.4 | 47.9 | 63.8 |
| | 0.8 | 70.6 | 80.0 | 75.0 | 56.5 |
| | 0.825 | 81.4 | 71.5 | **76.1** | 50.7 |
| | 0.85 | 87.4 | 61.6 | 72.2 | 43.8 |
| RepVGG-A1 | 0.75 | 58.7 | 83.7 | 69.0 | 58.9 |
| | 0.8 | 82.5 | 70.3 | **75.9** | 49.9 |
| | 0.825 | 87.0 | 61.2 | 71.9 | 43.4 |
| | 0.85 | 90.3 | 52.0 | 66.0 | 36.9 |
| RepVGG-A1 + P | 0.75 | 60.8 | 86.4 | 71.4 | 61.4 |
| | 0.8 | 85.3 | 69.2 | **76.4** | 49.8 |
| | 0.825 | 90.2 | 57.4 | 70.1 | 41.6 |
| | 0.85 | 93.0 | 43.1 | 58.9 | 31.4 |

Table 2 shows the results of SEResNet [12] and RepVGGA1 [10] as the encoder under the same condition, where P [16] means penalization. It does have improvement at F1 score when we apply penalization on the multi-head attention pooling layer.

### 5.5. Muti-language training

Table 3: Test results on Cantonese with multi-language trainset

| Testset Ⅱ | Threshold | Precision | Recall | F1 | MAO |
|---|---|---|---|---|---|
| 2k RepVGGA1 | 0.75 | 11.9 | 90.2 | 21.1 | 59.3 |
| | 0.8 | 27.9 | 77.0 | 41.0 | 51.0 |
| | 0.825 | 35.0 | 62.8 | **45.0** | 41.6 |
| | 0.85 | 41.1 | 44.9 | 42.9 | 29.5 |
| 4k RepVGGA1 | 0.75 | 10.5 | 93.3 | 18.8 | 61.2 |
| | 0.8 | 24.1 | 83.6 | 37.4 | 55.2 |

|  | 0.825 | 32.0 | 74.5 | 44.8 | 49.5 |
|  | 0.85 | 38.4 | 59.7 | **46.7** | 39.8 |

Table 3 shows it will enhance the performance on Cantonese if we add 2000 more English words in our training set.

### 5.6. Competiton

Table 4: Ablation experiments on Mandarin test set

| RepVGGA1-deploy retrained | Precision | Recall | F1 | MAO |
|---|---|---|---|---|
| √ nms/norm/comp | 75.7 | 76.0 | 75.9 | 53.9 |
| √ norm/comp | 75.2 | 76.2 | 75.7 | 54.1 |
| √ comp | 75.2 | 76.2 | 75.7 | 54.1 |
| × nms/norm/comp | 77.6 | 69.3 | 73.2 | 49.1 |
| SE-ResNet |  |  |  |  |
| √ nms/norm/comp | 81.4 | 71.5 | 76.2 | 50.7 |
| √ norm/comp | 81.0 | 71.6 | 76.0 | 50.8 |
| √ comp | 81.0 | 71.6 | 76.0 | 50.8 |
| × nms/norm/comp | 90.1 | 54.7 | 68.0 | 39.0 |

Table 4 shows that involving competition among words could enhance performance when testing, where √ donates using the strategy and × donates not using the strategy. And NMS operation will also enhance the precision as well as F1 score. Norm maybe useful under some conditions. In this set of comparisons, SEResNet [12] is superior to RepVGG [10] because of data enhancement.

### 5.7. Inference method

Table 5: Contrast test results with different inference methods

| Testset | Threshold | Precision | Recall | F1 | MAO |
|---|---|---|---|---|---|
| 160① | 0.75 | 60.8 | 86.4 | 71.4 | **61.4** |
|  | 0.8 | 85.3 | 69.2 | 76.4 | 49.8 |
|  | 0.825 | 90.2 | 57.4 | 70.1 | 41.6 |
|  | 0.85 | 93.0 | 43.1 | 58.9 | 31.4 |
| 296① | 0.75 | 40.2 | 90.9 | 55.7 | **63.9** |
|  | 0.8 | 75.7 | 76.0 | **75.9** | 53.9 |
|  | 0.825 | 86.5 | 62.7 | 72.7 | 44.7 |
|  | 0.85 | 91.1 | 46.8 | 61.9 | 33.6 |
| 296② | 0.75 | 41.1 | 88.2 | 56.1 | **62.1** |
|  | 0.8 | 76.5 | 73.7 | **75.0** | 52.3 |
|  | 0.825 | 86.9 | 60.8 | 71.5 | 43.5 |
|  | 0.85 | 91.2 | 45.2 | 60.4 | 32.5 |
| 160② | 0.75 | 56.3 | 86.8 | 68.3 | **61.6** |
|  | 0.8 | 82.5 | 71.4 | **76.6** | 51.2 |
|  | 0.825 | 89.2 | 57.8 | 70.1 | 41.8 |
|  | 0.85 | 93.2 | 41.2 | 57.1 | 30.0 |

Table 5 shows the experiment results on accelerating inference speed base on RepVGGA1 [10] model, where ① and ② below stand for inference with one single model and separate model, which means it could get comparable results in different inference schemes after we finetune the model, 296 and 160 here stand for the input size of T fbank frames. Finally, the last row shows the model trained with 160 frames as input as well as enhanced the number of parameters in the embedding layer.

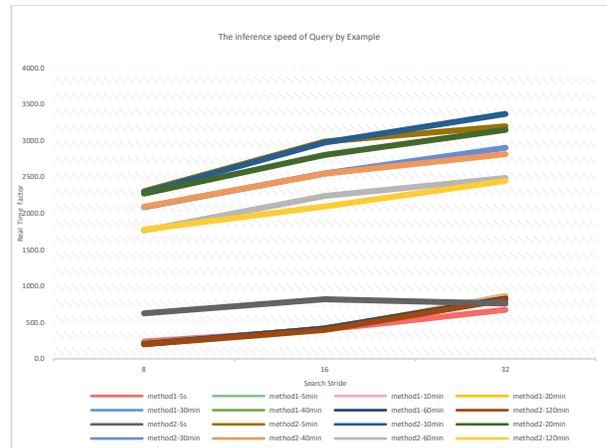

Figure 8: Contrast speed results with different inference methods

Fig. 8 above shows accelerate ratio under different conditions, where we could see that the longer the audio time $F_H$ and the shorter the time stride $s_t$, the better the effect of the separate model inference scheme compared with one single model.

### 6. CONCLUSIONS

In this paper, we propose a novel fast QbE speech search method based on separable models in order to enhance information utilization as well as speed. Moreover, we design a model test scheme in detail, and put forward the evaluation index for the scheme. We think that our work proves that the pre-trained parameters are meaningful when we transfer to the same task with different inference methods. It may be extended to other tasks that have sequential inference as well as information aggregation such as Speaker Clustering.